# Is there a Trojan! : Literature survey and critical evaluation of the latest ML based modern Intrusion Detection Systems in IoT environments

**Vishal Karanam**
University of Southern California, Los Angeles, CA

## Abstract

IoT as a domain has grown so much in the last few years that it rivals that of the mobile network environments in terms of data volumes as well as cybersecurity threats. The confidentiality and privacy of data within IoT environments have become very important areas of security research within the last few years. More and more security experts are interested in designing robust IDS systems to protect IoT environments as a supplement to the more traditional security methods. Given that IoT devices are resource-constrained and have a heterogeneous protocol stack, most traditional intrusion detection approaches don't work well within these schematic boundaries. This has led security researchers to innovate at the intersection of Machine Learning and IDS to solve the shortcomings of non-learning based IDS systems in the IoT ecosystem.

Despite various ML algorithms already having high accuracy with IoT datasets, we can see a lack of sufficient production grade models. This survey paper details a comprehensive summary of the latest learning-based approaches used in IoT intrusion detection systems, and conducts a through critical review of these systems, potential pitfalls in ML pipelines, challenges from an ML perspective and discusses future research scope, and recommendations.

## 1 Introduction

Internet of Things (IoT) encompasses tools/devices with sensors, computational power, software, etc, that form an interconnected system that can exchange information over the broader internet or other communications channels and with themselves. IoT as an ecosystem has grown manifold over the years with smart devices, applications in healthcare, and much more domains. The IoT market cap is valued at more than 400 Billion today and around 13.5 Billion devices are estimated to be a part of the IoT global ecosystem. With the ever-burgeoning malicious adversary attacks, spyware, and Trojans inundating the IoT network perimeters, there's been an ever-increasing interest in using Machine Learning techniques in conjunction with the fundamental concepts of computer security to tackle the problem of intrusion detection (1). Advancements in the fields of Machine Learning and Deep Learning have led cybersecurity researchers to dip their toes in the intersection of Machine Learning, IoT, and Intrusion Detection.

The first half of the paper deals with what an intrusion detection system is, how Intrusion detection systems in IoT environments differ from the traditional ones, the generic architecture of an IDS, and different approaches to classify the IDS based on learning approach and datasets. Then next part is a comprehensive discussion on few of the Machine learning based approaches in this domain.

Despite several breakthroughs and great potential in unlocking security research in IDS design, machine learning usage is not devoid of pitfalls that undermine model accuracy/performance in real scenarios due to training data selection, hyperparameter tuning, over/under-sampling, no rigid baselines for the model estimate, benchmark, etc.

These miscalculations while applying Machine learning pipeline workflows to ML-based IDS potentially render them unsuitable for practical deployment in real networks. The second half of the paper tries to evaluate the ML-based approaches and look at some of the deficiencies in using ML. The latter sections use the classification, analysis, and pitfall identification approaches proposed in the paper by Arp et al. (2) to evaluate the deficiencies in IDS for IoT environments using the latest IoT-relevant datasets.

The last part of the paper looks at the challenges involved, as there's still a lack of standardization on which learning-based approaches or datasets are ideal for a generic state-of-the-art IoT IDS. Furthermore, the lack of many production-grade IoT models in this space, despite multiple ML models having > 99% accuracy tells us that there's a gap to be bridged in evaluating the models more robustly and utilizing the up-to-date datasets in training our models.

**This paper's contribution can be summarized as:**

- Classifying the IoT IDS on the basis of Learning methods and datasets used for evaluation. **[Section 3]**
- Discussing the latest works which exclusively use IoT attack trace relevant datasets for evaluation. (Not older datasets (or) Non IoT) **[Section 3]**
- Critical evaluation of the Machine Learning approaches for the latest works. **[Section 4]**
- Recommendations and further research on this topic for security researchers utilizing ML.**[Section 5 and 7]**

## 2 What is an Intrusion Detection System

An IDS is a tool that helps detect Trojans/spyware/malicious traffic inside a system or a network. Our focus in this paper is on NIDS (Network based) and HIDS (Host-based). Often in the real world, it's best to use a combination of both in a hybrid sense and also a distributed IDS, which is scalable, available, and fault tolerant. An IDS designed for IoT environments must be able to scan data packets in network layers and stacks, generate responses in real-time, and be adaptable to different tech stacks in the IoT environment. An IDS that is designed for IoT-based environments should operate under stringent conditions of low processing capability, fast response times, and high-volume data processing.

The generic approach to security is to have the system engulfed in security mechanisms, such as authentication frameworks, encryption, firewalls, VPN's, etc so that they form a capsule around the system. But more often than not these techniques are vulnerable and have loopholes. Hence modern distributed networks need IDS as a complement to the existing security infrastructure.

### 2.1 Why do we need separate IDS for IoT ecosystems

The IDS deployed in an IoT environment in the network perimeters must be able to process, analyze and respond to data packets in different IoT network layers, sometimes in a heterogeneous network with multiple different network stacks. It must also be operational under more restricted conditions than a traditional IDS such as it must be functional in sensors of low processing capability, must have low latency and faster response time with limited hardware(3). An IoT environment consists of multiple different IoT devices interacting with each other and exchanging packets of information and data. IoT is a heterogeneous environment, that generates high-dimensional, multi-modal, and temporal data. Using standard ML and deep learning models for either training or validation on this data requires a lot of computational power. So, one approach which was looked by researchers is to leverage the power of the cloud for parallel processing, storage and availability. Standardization and synchronization between various clouds is a challenge. Since IoT data is bound by several privacy rules, traditional IDS which leverage cloud to store data in other jurisdictions can't be fully applied. In addition to that, latency issues and maintaining the security of the cloud itself posed a major challenge to researchers (4; 5).

There are many security and privacy challenges in IoT systems which are different from traditional IDS. Challenges like, hardware corruption, chip and processing power in the physical layer to distributed DoS attacks, gateway attacks, and packet sniffing in the networking layer. The application layer is prone to code attacks, code vulnerabilities like injection, scripting bugs, configurations, broken accessibility protocols etc (6).

Liu et al(7) describe the security analysis from an IoT perspective, and propose a unified framework for the low capability limits of IoT devices. Various threats can attack various layers of the IoT ecosystem as seen above, so Liu et al(7) discusses the information flow in the network and potential security and privacy problems.

Their paper classifies the security challenges into categories which are:

Authentication and physical threats : A wireless sensor or an RFID tag may dubiously claim it being in a location different from its real existence.

Integrity : IoT data is easily spoofable and tamperable.

Confidentiality and Privacy: Since IoT devices lack processing power, this presents a challenge to ensure confidentiality, as it introduces barriers to apply standard encryption and key exchange algorithms. Furthermore, since IoT devices are mostly wireless, snooping over the wireless communications is another challenge.

For these reasons, conventional IDS may not be fully suitable for IoT environments. IoT security also needs to continuously evolve in consideration of the ever-changing threats, malicious agents attacking the very core stability model of the IDS software, and passive analysis of network data packets for the long term stability of the system. This led security experts to further pursue offline machine learning, which trains the model with pre-existing datasets and uses the models in production directly to detect threats, thus saving the time on sniffing for packets real-time, which is an expensive operation. The machine learning models would also be updated regularly to detect newer threats.

### 2.1.1 Generic Architecture of Intrusion Detection Systems

The types of attacks troubling IoT systems are Dos attacks (Syn flood), Spoofing,( Ipsweep, portsweep, nmap), Compromise (R2l, U2R), Virus, worms ,Trojans etc.

The operations of an IDS can be divided into three stages. The preliminary stage is the data collection phase. Typically in IoT environments this is a sensor, or a network analyzer. Data is also collected from logs, kernel application dumps, audit trails, protocol level sniffers, gateway analysers, Simple Network Management Protocol (SNMP) , application session logs, wireless networks etc.

This stage relies on network-based or host-based sensors. The second stage is the analysis stage, which relies on feature extraction methods or pattern identification methods. It consists of an analysis engine that collates the data and packets collected from the monitoring stage and stores in the knowledge databases. This data is refined by knowledge base of attacks and their signatures, filtered data, data profiles, etc. IDS systems also have a Config setting, which tells about the current state of an IDS and whether itself is available and ready(8).

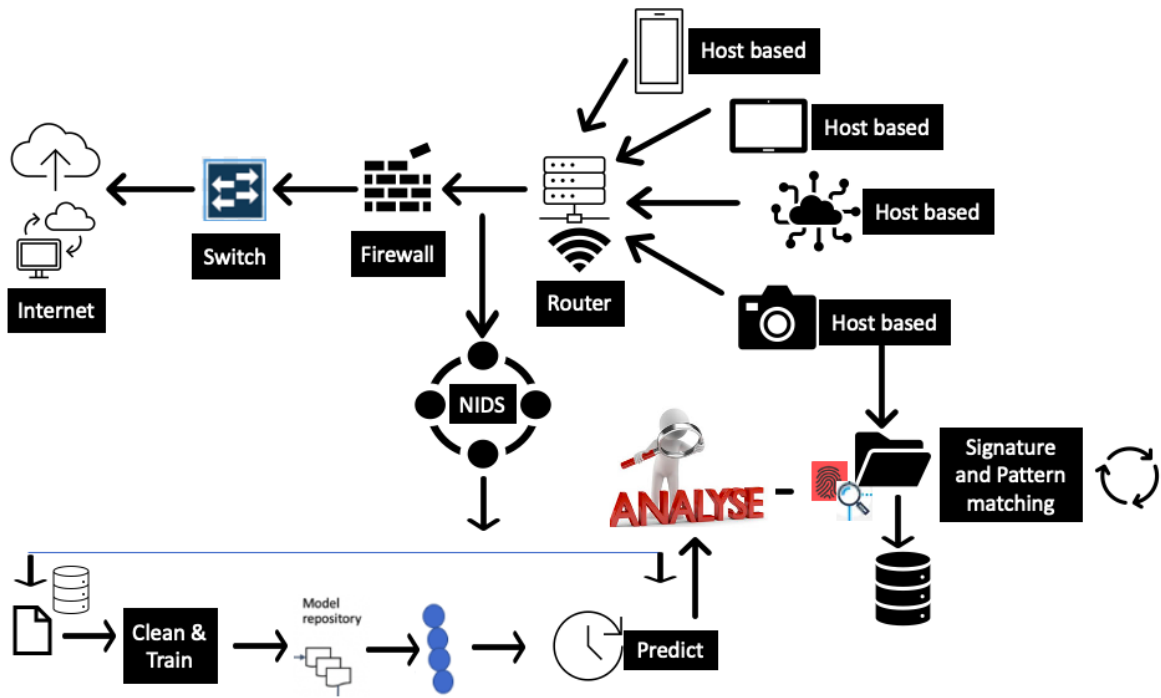

Figure 1: Generic Architecture of an IDS in IoT Ecosystem

The final stage is the detection stage, which relies on anomaly or misuse intrusion detection. An IDS captures a copy of the data traffic in an information system and then analyzes this copy to detect potentially harmful activities. Post this stage the IDS works to respond to the attack either autonomously or through human agents.

An ideal IDS should have good prediction performance, low false negative rate, good processing speed in resource constraint IoT environments, good ROC-AUC curve, distributed in nature and fault tolerant.

A network-based intrusion detection system (NIDS) sniffs network traffic packets to detect intrusions and malicious attacks (9). A NIDS can be either a software-based system or a hardware-based system. For example, Snort (10) NIDS is a software-based NIDS. The operational structure of a NIDS and its location in the network are shown in the Figure 1.

IDS depends on statistics, pattern matching and ML algorithms for implementing the various sub stages of intrusion detection. IDS algorithms can be classified as signature based or anomaly based.

# 3 Detection Techniques used in Intrusion Detection Systems

IoT environments are susceptible to many kinds of attacks. The surface area and perimeter of exposure in IoT environments leads to attacks such as Man in the middle, DDoS, Service scan, Zero-day attacks, data exfiltration and key-logging etc. Two of the most common kinds of detection algorithms used in IDS are Misuse-based intrusion detection and Anomaly based.

A misuse-based intrusion detection technique uses a knowledge history of known signatures and patterns of malicious codes and intrusions to detect well-known attacks. Some known constrains are Network packet overload, the high cost of signature matching, memory constraints wrt storage , continuous database key-value updates(8).

In an anomaly-based intrusion detection technique, a perfectly stable and normal data pattern is created based on data from valid users and is then compared against current data patterns to detect anomalies or abnormal behaviour(11).

Some other approaches used in Intrusion detection are Data discovery approach which uses summarizing and scrapping techniques, commonly used in data mining to extract knowledge from a massive traffic and network datasets. This knowledge can be combined with algorithms to predict behavior of data from users or networks. Other classifications also include rule-model-based , formalism based, specification based etc. Several supervised, unsupervised, transfer learning, pattern detection models have also been suggested for various IDS cases .

Security researchers typically look at 2 broad classifications of IDS in IoT environments. One where the IDS solutions utilize machine learning-based techniques, and one where they don't. The focus in this paper is on the former, but as a precursor let us briefly take a look at the latter.

### 3.0.1 Classification based on Non-Learning approaches

Raza et al.(12) proposed SVELTE, the first IDS of its kind for IoT environments. SVELTE is a real time IDS that combines both signature and anomaly detection methodologies and focuses on spoofing and sinkhole attacks targeting IoT devices using IPv6. SVELTE had a weakness as it was susceptible to DoS attacks. Shen et al.(13) proposed an device fingerprint hybrid approach based IDS for ICS networks with a memory database to cordon the security perimeter. This could be extrapolated to IoT environments which are similar to the ICS network in terms of processing, threats, storage, computation power etc.

Jun and Chi(14) proposed a CEP event-processing IDS architecture that uses a rule based approach and performs particularly well in detection of Trojans in real time.

Krimmling et al.(15) proposed a NIDS framework for smart cities transport networks that use CoAP (Constrained Application Protocol useful) . A CoAP is a specialized transfer protocol used in resource constrained IoT networks.

Several state of the art IDS have been proposed based on 6LoWPAN in IoT networks. Kasinathan et al. (16) proposed a Denial of service attack detection architecture which was a very novel approach for 6LoWPAN in 2013. Surendar et al. (17) proposed a constraint based IDS and response navigation system for IoT networks. Le et al (18) also proposed a specification IDS to defend attacks such as sinkhole, local repair and other RPL topology attacks.

Midi et al.(19) proposed Kalis which is a combination of signature and anomaly-based IDS which is adaptable to network features. This is very effective as it's protocol independent.

### 3.0.2 IDS for IoT environments based on Machine Learning techniques

Sniffing for real time data in huge networks and detecting for threats is a computationally expensive process. Instead models can be trained on data, both benign and non-benign, captured apriori and then deployed in real time to detect malicious actors. For this the training data and the model itself must be continually updated and trained with the latest malicious actors inundating the IoT space.

Recently there has been a lot of focus on applying machine learning and deep learning models to detect security threats and vulnerabilities in the IoT space. After machine learning standing out as the go to approach in spam detection, credit card fraud detection, text classification and anomaly detection space, many security researchers have put forward the case to use ML/DL models in IDS systems. Based on the results and accuracy of the various state of the art methods, ML/DL seems to be a good choice for architecting anomaly based IDS. The Machine Learning Taxonomy is presented in this paper to classify the IDS with the algorithm used in modelling.

The top level classification for IDS based on learning methods is :

1. Supervised
2. Unsupervised
3. Semi-Supervised
4. Deep Learning
5. Mix and Match(Hybrid)

**Interesting and state of the art approaches :** Several papers have been looked at and analyzed, belonging to these categories. The next subsections classify them on the type of machine learning technique and takes a look at them in depth. The paper by Hodo et al.(20) which uses an Artificial Neural network is particularly interesting. It combines transfer learning and stochastic learning methods to analyze the IoT Networks IP stack. Furthermore, their paper also simulates the working of their model in a real simulated environment where it bears the brunt of a few DoS attacks. The model also has interesting applications in other resource constraint environments like ICS. Another fascinating paper is the one by Koroniotis et al.(21) the original creators of the Bot-IoT dataset. They laid the foundation for IoT specific datasets to be used for better accuracy and go in depth covering the various kinds of attacks affecting the IoT systems. Additionally, they make the case for ML models to be tailor made to the kinds of attacks rather than over generalization. Lastly, despite deep learning's disadvantages with respect to performance and time complexity, Abdalgawad et al's.(22) auto encoder and Generative deep learning models are a breakthrough. Their models do a great job in detecting attacks like distributed denial of service, and various botnets like Mirai, Okiruk and Torii.

**Supervised Machine Learning :**

Khraisat, et al. (23) (24)proposed an ensemble Hybrid IDS by combining a C5.0 Decision Tree classifier and a One-Class SVM. C5.0 DT classifier is used to detect well know intrusion. One-Class SVM classifier is used to detect a new attack. This approach classifies both known and Zero-day attacks with good accuracy. This ensemble method has an accuracy of 97.40 on the ADFA dataset. Nobakht et al.(25) used a combo of Logistic Regression and SVM for anomaly detection in Real IoT Hue lights. Their method IoT-IDM uses both an SDN and Machine Learning as detection strategies.

Hodo et al.(20) proposed an ANN with 3-layer Multi-Layer Perceptron (MLP) approach for IDS design with an emphasis to combat DDoS attacks in the IoT ecosystems. The hidden layers of the ANN used a unipolar sigmoid transfer function and a stochastic learning algorithm with a mean square error function was used. Their analysis is built on Internet packet traces and the validation of the proposed method was done on a simulated IoT network with an accuracy of 99.4 percent. The simulated IoT network comprised of 4 clients and one server. The malicious simulated adversaries targeted the server node by sending more than 10 million packets ultimately making it go to a failed state.

Thaseen et al(26) suggested an IDS using correlation-based attribute selection (sorts features based on the correlation coefficient) and ANN. The suggested approach was tested against the NSL-KDD and UNSW-NB datasets. Ashraf et al (27) developed a statistical learning-based IDS which aims to protect the network perimeters of smart cities. The paper investigates attacks such as Mirai and bashlite which compromise nodes in smart cities through DDoS attacks. The paper explores IoTBoT-IDS which uses Beta Mixture Model (BMM) and a Correntropy model that learns the normal behaviour and flags any abnormal behaviour and classifies it as a security threat. The results were evaluated against 3 realistic IoT environment datasets.

Nguyen, Ninh and Hung's paper (28) on MidSiot( A Multistage Intrusion Detection System for Internet of Things) presents a new state of the art distributed IDS for IoT use cases in both resource constraint and heterogeneous hardware specification stacks. This 3 stage approach has been evaluated against IoTID20, CIC-IDS-2017, and BOT-IoT datasets with 99.68% accuracy.

Ferrag et al. (29) proposed RDTIDS: Rules and Decision Tree-Based Intrusion Detection System for Internet-of-Things Networks which uses multiple approaches (DT and rule based): REP Tree, JRip algorithm and Forest PA. It uses 3 classifiers where the result of the first 2 classifiers which happen simultaneously are used by the 3rd one. The model is tested against the real traffic data set CIC-IDS-2017 and BoT-IoT and performs with an accuracy of >96%.

Soe YN, Feng Y, Santosa PI, Hartanto R, Sakurai K (30)suggested a supervised approach CST-GR algorithm implemented on raspberry pi. This approach is unique in the sense that it's very light weight and very fast. CST-GR employs a feature selection criterion such that only the most important and necessary features are selected which betters it's time and space complexity. This approach has an accuracy (TPR) of 99.4% when J48 and RF being used as the classifier and CST-GR does the feature selection.

A. Alhowaide, I. Alsmadi, J. Tang (31) suggest emsemble feature selection methods, PCA and Random Forest for dimensionality reduction. The methodology proposed uses BoTNeTIoT-L01, the latest IoT dataset, containing real IoT data traffic sniffed through Wireshark and two Botnet attacks (Mirai and Gafgyt).

Aldhaheri et al. (32) present a system which models a neural network similar to the human body and achieves an accuracy of 98.73 on the IoT-BoT dataset. Their model tries to reduce false positives and performs well in comparison to other supervised methods.

Koroniotis et al.(21) the original founders of the Bot-IoT dataset, the most recent and the largest publicly available dataset with IoT Trace data, fit multiple ML models, created a usable test simulation environment, reported the accuracy and performance.

**Semi Supervised Machine Learning :**

Supervised ML has a good accuracy in general but does a poor job with zero-day attacks whereas Unsupervised ML has lower accuracy but has good detection capability on unknown/zero-day attacks. Hence, Rathore and Park (33) proposed a semi-supervised Fuzzy C-Means approach using the supervised and unsupervised ML for labelled and unlabelled inputs. This distributed Network IDS utilizes a new paradigm called fog computing and works in distributed and low latency resource constrained IoT ecosystems.

**Unsupervised Machine Learning :**

Lopez-Martin et al.(34) proposed an unsupervised anomaly NIDS for IoT based on Conditional Variational AutoEncoder (CVAE). Their method is unique due to its ability to carry out feature reconstruction, i.e., it can retrieve missing features from incomplete training datasets. Authors claimed to achieve 99% accuracy.

Apostol et al.(35) proposed an Anomaly Detection model using Unsupervised Deep Learning. They use a deep auto encoder model with ReLU for the hidden layers. This method when evaluated with the Bot-IoT traffic from IoT devices and cloud resulted in 99.7 %for accuracy and 99% for precision.

**Deep Learning :**

Khan, M.A. (36) suggested a hybrid convolutional recurrent neural network to classify cyberattacks. The guiding principle is leveraging the fact that the CNN better models the spatial features while the RNN models the temporal features and use the benefits of both Anomaly/Signature based approaches. The proposed HCRNNIDS has an accuracy of up to 97.75% for the CSE-CIC-IDS2018 dataset with 10-fold cross-validation.

Abdalgawad et al.(22) used generative deep learning methods like Adversarial Autoencoders (AAE) and Bidirectional Generative Adversarial Networks (BiGAN) to successfully classify IoT threats. The results were evaluated using the IoT-23 dataset based on a limited set of IoT devices like Somfy door lock, Philips Hue and Amazon Echo. Their premise was to use generative deep learning models to thwart DDoS attacks and BotNets like Mirai and Okiruk. Empirically their models when tested on IoT-23 got an F1-Score of 0.99 and were also able to predict zero-day attacks with an F-score between 0.85 and 1. This is one of the few approaches which predicts zero-day attacks with good accuracy.

Ullah and Mahmoud (37) presented a CNN as an alternative to anomaly detection to build a multi class classification model. Classification is done through 1D,2D,3D CNN models and transfer learning is used to build multi class models. The proposed CNN is validated using latest IoT ecosystem datasets like BoT-IoT, IoT Network Intrusion, MQTT-IoT-IDS2020, and IoT-23 intrusion detection datasets with a comparable accuracy to the state of the art approaches. The base detection rates of the 1D, 2D and 3D models are 99.74, 99.42%, and 99.03% respectively for BoT-IoT, MQTT-IoT-IDS2020, IoT-23, and IoT-DS-2 datasets.

**Mix and Match Approaches :**

Pajouh et al.(38) presented an anomaly IDS built with Two-layer Dimension Reduction and Two-tier Classification (TDTC) for IoT Backbone with the First layer unsupervised and the second layer supervised.

Alani and Miri(39) proposed a practical recursive feature selection approach and forbids relying on traditional feature reduction algorithms like PCA or SVD as they can have a huge time penalty in IoT environments and create a processing bottleneck. TON_IoT, IoT-ID, and IoT-23 datasets were used to validate the hypothesis and produced an accuracy of 99% in validation.

Abdulaziz et al. (40) proposed a hybrid model which uses Convolution networks and a transient search technique to balance between different phases of operation. The unique thing about their work is that they test their proposal on both the older NSL KDD datasets and also the new BoT-IoT dataset and provide a more nuanced estimate of performance.

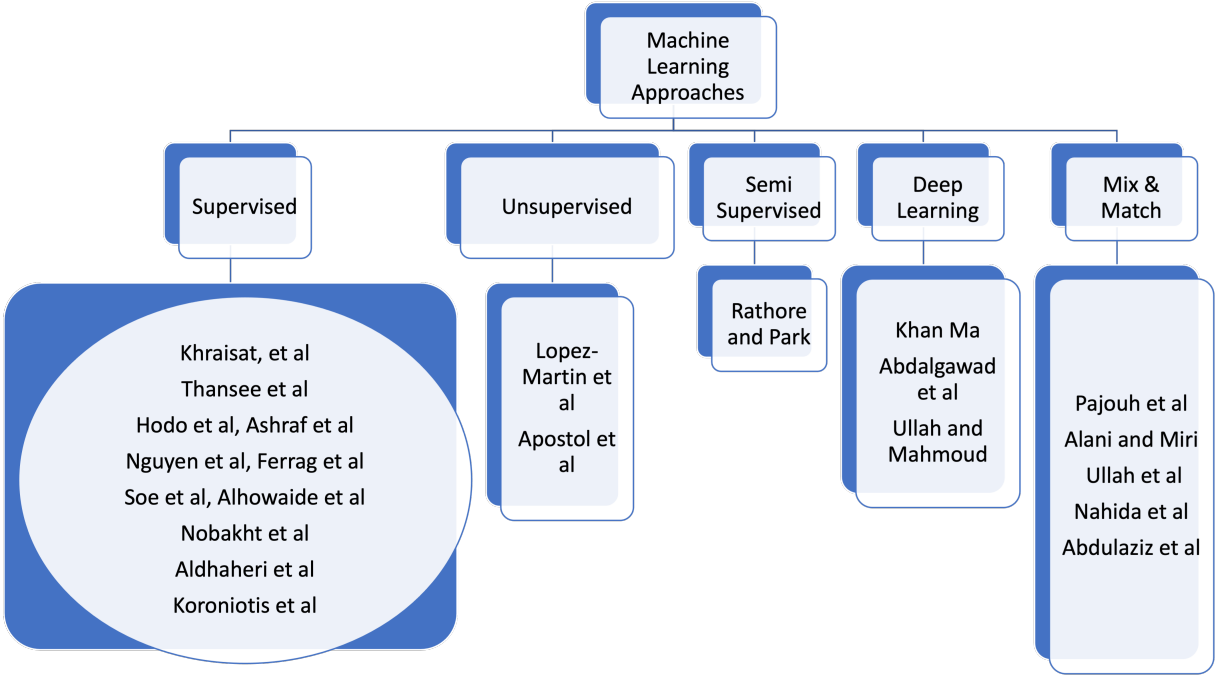

Figure 2: Classification of an IDS based on the ML approach

Nahida et al(41). have suggested using a Bi-LSTM after evaluating against both newer datasets like IoT Botnet and IoTID20 and older KDD datasets. As per their experimentation, Bi-LSTM performs the best and they have also suggested optimizations for using it in low power environments.

Ullah et al.(42) use a hybrid convolutional neural network (CNN) and gated recurrent unit (GRU) to detect and classify binary and multiclass IoT network data. The multiclass classifier has 2 CNN and 2 GRU layers while the binary one has 1 each. The hypothesis is evaluated using BoT-IoT, IoT Network Intrusion, MQTT-IoT-IDS2020, and IoT-23 datasets. Vitorini et al(43) and Albulayhi et al. (44) also put forth various approaches, taxonomies and guidances for intrusion detection techniques for IoT environments.

### 3.0.3 Taxonomy of learning-based IDS with respect to Datasets used for training and testing.

The second Taxonomy proposed in this paper is to classify the learning-based IDS models wrt the dataset used for testing.

Selecting the right datasets for training machine learning models is very important to get the desired results. There is a severe drought of available literature in this area due to data complexity, privacy of data and the difficulty of capturing and segregating data to make it explainable and useful (45). A model can give us a very high accuracy on a particular dataset but that would be of no use to us, if the dataset itself is not a representative of the problem being tackled.

This paper only considers models which are tested with the latest IoT trace datasets. The models trained on old datasets, while widely referenced are no longer trusted to predict new attacks in the IoT ecosystem. A few publicly available datasets used for Intrusion Detection in the IoT space are KDD(46), NSL-KDD (47), UNSW's dataset(48), BoT-IoT, MQTT-IoT-IDS2020 (49),CIC datasets (50), ToN-IoT & IoT-23(51)(52), BoTNeTIoT-L01. The KDD, NSL-KDD and UNSW's dataset are prominently featured and used by many IDS systems, but the issues with using them for training models are recency of data, lack the IoT traffic, unbalanced class wise and need sampling techniques to avoid biases, non-representative of the feature set(extrapolation) etc. This paper looks at recent datasets which are more relevant to the IoT ecosystem primarily BoT-IoT, MQTT-IoT-IDS2020,CIC datasets, IoT-23, BoTNeTIoT-L01 and study the ML approaches which utilize them.

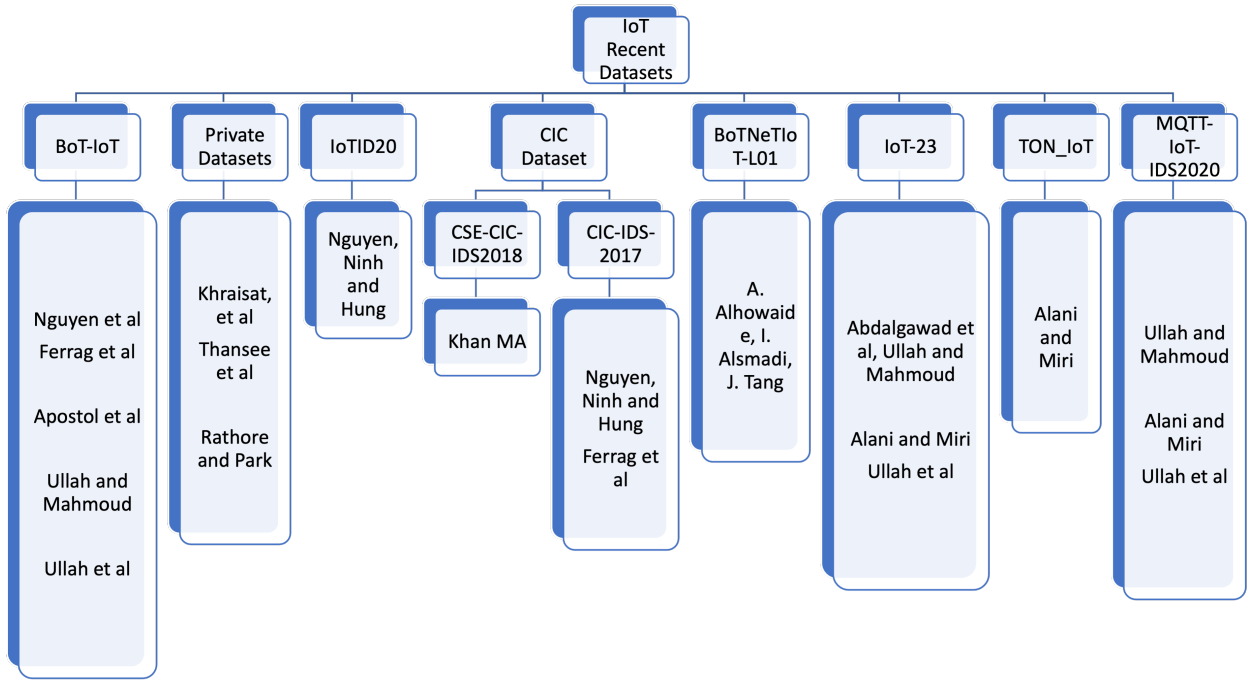

Figure 3: Classification of an IDS based on the Dataset used

## 4 Pitfalls of Intrusion Detection Research involving ML

Several advancements have been made combining ML with Security research and there have been significant breakthroughs. This development has influenced several related areas of computer security, such as work on learning-based security systems, malware/trojan detection, vulnerability discovery, and code analysis.

Sommer et al (53)have made an attempt to identify pitfalls and issues with using ML for intrusion detection systems and practical considerations.

This research in ML deficiencies has seen tremendous interest, and has been extended more recently to other domains, such as malware analysis and website fingerprinting. The highly cited paper(2) Do's and Don'ts of Machine learning published in USENIX-22 has provided a taxonomy to classify the common pitfalls of Machine Learning workflows in the security domain. Arp et al.(2) discuss the common pitfalls, machine learning approaches are prone to be exposed to in the security domain and suggest recommendations. Arp et al.(2) surveyed 30 papers from A* security conferences over the years and from their research show that these gaps/pitfalls are spread out in the security domain.

Extending the line of work by Arp et al, this paper tries to discuss the generic pitfalls related to machine learning security in IDS for IoT Environments. This part of the paper plays the devil's advocate and critically evaluate the various learning-based approaches through the Do's and Don'ts lens.

The general ML pipeline can be divided into 4 key steps(2). They are Data collection and labeling, Model design, evaluating the designed Model and making the model live in production. The Intrusion detection approaches in IoT ecosystems which use learning-based methods also follow the similar sequence of actions. In each of these steps the Figure 4 identifies potential fallacies which can be easily overlooked during the design and development stage.

Various studies have been looked at to understand what parameters need to taken into consideration for the IDS in IoT ecosystem case to make the critical evaluation meaningful(54; 55; 56; 57; 2). These works also help how to identify the pitfalls and what criteria must satisfy for the deficiency to be deemed as present.

**Base line Parameters used for the taxonomy to critically evaluate issues associated with ML/DL approaches :**

1. **Sampling bias** : The distribution of the data used to train the ML models is different from the real-word data. This happens when the variety/complexity in the real-world data is not accounted for during data collection.

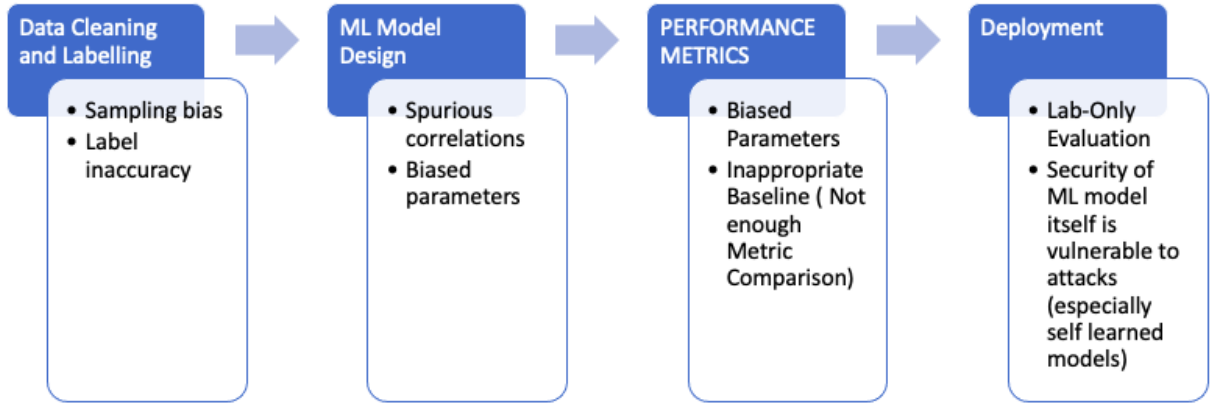

Figure 4: General Machine Learning Pipeline Workflow

2. **Label Inaccuracy** : Supervised methods rely heavily on label information in data. As a result, inaccurate and erroneous labelling adversely affects model performance.
3. **Spurious Correlations** : ML models are prone to overfit to training data and learn patterns that do not generalise to real world data. Consequently, the learning model adapts to these spurious patterns/ artifacts instead of solving the actual task.
4. **Inappropriate Baseline** : When the proposed ML models are evaluated without, or with limited, baseline methods it is not possible to demonstrate improvements against the state of the art.
5. **Biased Parameters** : When the learning parameters of the models are not tuned using a validation set and instead only optimised over the training set. These parameters may not give the best performance during test time.
6. **Not enough Metric Comparison** : The performance metrics used to evaluate the ML models should be in-line with the application scenario. For instance, if the labels are imbalanced then accuracy is a poor metric as it does not give an accurate measure for model performance on the minority class. Instead, class wise precision and recall can be used.
7. **Lab-Only Evaluation** : Using sufficient compute resources, learning models can achieve good performance on data collected in laboratory setting. But it is difficult to say if this performance is transferable to real world setting. Thus it important to test the models in a real world setting and report observations.
8. **Inappropriate Threat Model (Security of ML model)** : The security of model itself isn't taken into account. The model is itself prone to be outdated if doesn't used the latest dataset or if it's not regularly update with current info on attacks in the IoT space. The attackers can implant various DoS attacks, evasion attacks and make the model redundant. If there is no discussion on this, then the table has an entry of present , stating a possible pitfall.

The table below is a summary after parsing the papers using the latest IoT datasets through the pitfall checks.

Table 1: Figure 5 results

| Field Value | Description |
| --- | --- |
| Present | The bias is present in the paper |
| Present* | The bias is partially present in the paper |
| Not Present | The bias is not present in the paper |
| Unknown | The presence of bias is unknown due to dataset being private |

**Inference from the Critical Evaluation:** This paper critically evaluated the IDS which utilize learning-based methods and summarized the findings in the below table. From the evaluation, it's learnt that sampling bias is more prevalent in the datasets which are public and label inaccuracy is present to some extent in most datasets. The private datasets/simulations/testing methods couldn't be evaluated due to their opaque nature.

| TYPE OF ML CLASSIFICATION | PAPER | Sampling Bias | Label Inaccuracy | Spurious Correlation | Biased Parameters | Inappropriate Baseline | Not enough Metric Coverage | Lab-Only Eval | Inappropriate Threat Model (Security of ML model) |
|---|---|---|---|---|---|---|---|---|---|
| | | DATA COLLECTION AND LABELING | | | ML MODEL DESIGN | | PERFORMANCE METRICS | | PRODUCTION GRADE |
| SUPERVISED | Khraisat, et al | Unknown | Unknown | Not Present | Not Present | Present | Present* | Present | Unknown |
| | Thansee et al | Unknown | Unknown | Not Present | Not Present | Not Present | Present* | Present | Unknown |
| | Hodo et al | Present | Present | Not Present | Not Present | Present | Present* | Not Present | Present |
| | Ashraf et al | Present | Present | Not Present | Not Present | Not Present | Present* | Not Present | Present |
| | Nguyen et al | Not Present | Present | Not Present | Present | Not Present | Present* | Present | Present |
| | Ferrag et al | Present | Present | Not Present | Present | Not Present | Present* | Present | Present |
| | Soe et al | Not Present | Present | Not Present | Present | Not Present | Present* | Not Present | Present |
| | Alhowaide et al | Present | Present | Not Present | Present | Not Present | Present* | Present | Present |
| | Nobakth et al | Unknown | Unknown | Not Present | Not Present | Not Present | Present* | Unknown | Unknown |
| | Aldhaheri et al | Present | Present | Not Present | Not Present | Not Present | Present* | Present | Present |
| | Koroniotis et al | Present | Present | Not Present | Not Present | Not Present | Not Present | Not Present | Present |
| SEMI SUPERVISED | Rathore and Park | Unknown | Unknown | Not Present | Present | Not Present | Present* | Present | Present |
| UNSUPERVISED | Lopez-Martin et al | Present | Present | Not Present | Not Present | Not Present | Present* | Present | Present |
| | Apostol et al | Present | Present | Not Present | Not Present | Not Present | Present* | Present | Present |
| DEEP LEARNING | Khan Ma | Present | Present | Present | Not Present | Present | Present* | Present | Present |
| | Abdalgawad et al | Present | Present | Present | Not Present | Not Present | Present* | Present | Present |
| | Ullah and Mahmoud | Not Present | Present | Present | Not Present | Present | Present* | Present | Present |
| MIX & MATCH | Pajouh et al | Present | Present | Present | Not Present | Not Present | Present* | Present | Present |
| | Alani and Miri | Not Present | Present | Present | Not Present | Not Present | Present* | Present | Present |
| | Ullah et al | Not Present | Present | Present | Not Present | Present | Present* | Present | Present |
| | Nahida et al | Not Present | Present | Not Present | Not Present | Not Present | Present* | Present | Present |
| | Abdulaziz et al | Present | Present | Not Present | Not Present | Not Present | Present* | Present | Present |

Figure 5: Pitfalls Summary of the IDS in IoT using latest datasets

In deep learning-based approaches, raw data is given to the model as input and both extracting features and classifying the input are handled by the network. While this approach circumvents the need for manual effort in identifying the right features, they lack the explainability of the traditional supervised ML techniques like SVM and decision trees. Consequently, deep learning models are susceptible to learning spurious patterns and it is difficult to identify this patterns.

In terms of Model evaluation strategies, most papers fall short as there's either a lack of holistic comparison with multiple other models for different scenarios, a lack of comparison with non-learning based approaches, insufficient comparison amongst all kinds of ML approaches or no meaningful discussion at all. In terms of model's production deployment capabilities, most models fall short as they are only simulated in a lab specific environment and not an environment more close to the real world IoT ecosystems. This kind of evaluation is often overlooked with very few papers which even talked about a real world simulation. Models need to be tested and tried in a kind of IoT test-suite before being deployed in real world for more accurate performance. The model designed itself is vulnerable to attacks by adversaries. Some malicious hackers try to target the self-learning model with false data and mess up its algorithm. Some adversaries try to hack the model, create a Distributed DoS attack hampering its performance by overwhelming it, or learn its parameters while others can trick the model by training the attack itself to avoid the model's detection span. This can be done through various methods as some of the publicly available datasets are easily obtainable and the attacker can fit several ML approaches discussed and mask the attack to avoid getting detected by the IDS. The security of the ML model itself and continually updating the model(with latest attack knowledge) is very important and this an area which is not well discussed in the IDS systems surveyed.

**The purpose of the evaluation in this section is not to downplay any contributions, but to learn and understand the challenges in designing the systems and also to help security researchers to design more robust solutions in this space. The hope is to discuss the shortcomings, so that future research in this space is not prone to them**.

While the intrusion detection systems have great accuracy reported, there are quite a few things as discussed above, which were overlooked while estimating which might have led to overestimation (58).

These pitfalls can play spoilsport in analysis and result in overestimating the true accuracy, affect the design of the machine learning workflow, let researchers make incorrect assumptions and also cause failures when models are deployed in the real world.

# 5 Limitations and Challenges to overcome for learning-based Intrusion Detection in IoT ecosystems:

Several challenges need to be overcome to create a robust production grade IDS in IoT ecosystems. The datasets used by a lot of the learning-based approaches are older and stale datasets with respect to IoT traces, traffic and threats. Even though those approaches have high accuracy, their deployment on a large scale production level is questionable, given their lack of training on newer kinds of IoT traffic and attacks. The bottleneck is that there are very few publicly available datasets that cater to this domain.

The second challenge is collecting data points. Since IoT is a huge heterogeneous domain, collecting the data points and feature sets is a challenging and time consuming affair. Furthermore, a one size fits all collection of data is questionable and even if done the algorithms trained on them might be difficult to both explain and understand.

The third limitation which is stopping these approaches from attaining practicality is a lack of simulation in real IoT environments. Most approaches using ML, especially deep learning despite performing really well, are computationally expensive and given the resource constraint environments that IoT devices are, their performance in production falls short. The lack of publicly available real time IoT simulated environment for researchers to test is another bottleneck to further exploration. Unless researchers are able to test the models in close similarity to real ecosystems it becomes difficult to improve upon the shortcomings. The creation of a simulated environment itself is quite challenging and has many constraints with respect to scale, cost and maintenance.

The fourth limitation is the performance metrics. Most approaches discussed in this paper focus only on either accuracy, F statistics, TPR, FPR, ROC-AUC curve etc as baselines for model performance. But there needs to be an update to the metrics as the models need to be evaluated holistically with respect to accuracy, time taken to train, response time, computational complexity, latency etc.

# 6 Related Work

There have been many papers evaluating intrusion detection systems. This paper differs from them in the context that the current work evaluates those ML based IDS in IoT environments which use newer IoT relevant datasets. The older datasets like KDD99, NSL-KDD etc though have good amount of data, they are expired wrt IoT traffic and also don't have the kind of attacks penetrating IoT environments. The work in this paper also critically evaluates the ML models with respect to several parameters and the pipeline workflow itself, which is also another major differentiation point from other related works in this area.

The studies below in their recommendations, have also highlighted the issue with extrapolating models designed on older datasets to newer IoT environments. Several highly cited works (59; 60; 61) have been done by researchers at the intersection of security and ML and suggest explainable frameworks, methodologies, issues with using ML in IDS, classifications and concepts.

One of the earliest works of IDS classification was the highly cited paper of (Liao et al (62)) which reviews concepts related to IDS, IPS and a classification of IDS into signature based, anomaly-based or SPA.

Khraisat et al (23)discuss detection systems, techniques, datasets and challenges and also present novel IDS for IoT environments. The highly cited paper of Benkhelifa et al. (63)discusses the Practices and Challenges in IDS for IoT, while also critically suggesting the need for robustness, different research directions and critical recommendations.

Chaabouni et al (64) discuss in-depth the threats and challenges for IoT networks by evaluating, comparing and analyzing state-of-the-art IoT NIDS in terms of architecture, detection strategies, validation techniques & deployment modes. It also gives a thorough review of open source datasets, tools and sniffers.

The highly cited paper of Elrawy et al.(8) surveys Intrusion detection systems as a whole for IoT Networks. They have reviewed several summary papers and presented an in-depth taxonomy of various approaches to IDS and the general architecture of an IDS. Additionally, the work of both (64) and (8), looked at both non-learning based approaches as well as learning-based approaches. Furthermore, they suggested challenges to solving the IDS problem in IoT networks and have a detailed set of recommendations for further research on this topic. Several other works (65; 36)have analyzed IoT IDS from a ML/Deep learning perspective and suggest further deep dive into using ensemble methods.

H. Hindy et al (49)propose a new taxonomy and survey state of the art intrusion detection system design techniques, network threats and datasets particularly on the newer related data, while Yang et al.(66) and Zarpelao et al(67), also present a comprehensive survey on intrusion detection in internet of things. The more recent works particularly that of Khraisat and Alazab(68), extend taxonomy classifications, propose challenges related to IoT dataset collation, recent research on models to improve accuracy and results of IoT IDS and recommendations for further research.

Nahida et al.(41) have done a comprehensive evaluation of multiple ML techniques and also reviewed the existing methods and summarize them. They have provided multiple taxonomy classifications and used both the Older (KDD) datasets as well as the newer datasets like IoT Botnet and IoTID20. As per their experimentation Bi-LSTM performs the best amongst the deep learning techniques. AbdulAziz et al. (69) have proposed using a feature selection method with Aquila optimizer.

# 7 Recommendations and Future work

Since one size fits all approach is difficult to implement in IoT ecosystems, the recommendation is to use a majority voting strategy and combine learning-based IDS approaches with newer areas of research which are showing promise. Some early research in combining multiple domains of computing by Spadaccino et al.(70) have yielded promising results. Combining ML with edge computing and fog computing is one area which should be further explored.

Creating a reliable dataset for IoT use cases must be explored. With privacy challenges involved in using data, masking and non-attribution techniques could be evaluated to look at how the data could be collected and used for model training purposes.

More research on better data cleaning and sampling pertaining to IoT data points, holistic evaluation techniques to try to eliminate pitfalls and biases in data collections could go along a long way in pushing the models to a realistic scenario. Standardization and interoperability of IoT devices must be considered so that IDS research can be streamlined as well as data collection, processing and monitoring .(71)

Just having an IDS is not sufficient, which is often the case in IoT environments. Its of paramount importance to initiate a response in the fastest time. So there is a need to combine IDS and IRS for an effective security package. Additionally, models must be continually updated to keep them active to track new malicious threats. There are a lot of opportunities in this area. New emerging response systems with autonomous agents capable of responding based on AI is a fascinating area to be explored.

Dataset selection must follow guidelines like those researched by Torralba et al.(72) to avoid biases and sampling issues. More collaboration between ML and Security domains is needed to minimize drawbacks pertaining to model training, selection, testing, cross validation and model security.

Zhang et al.(73) have done novel work in using genetic algorithms and belief systems for intrusion detection and it has proven to be effective in a lot of use cases. It would be highly useful to try combining this approach with a non-learning based IDS and see its results in a simulated environment.

Additionally, The paper recommends exploring the untapped area of establishing model benchmark estimation holistically and not just with respect to accuracy, F-scores/TPR,FPR but also performance, time complexity and TTR (response time) for a more comprehensive and robust IDS design. Additionally, new research needs to carefully evaluate the ML pitfalls while designing newer models.

# 8 Conclusion

The IoT environment , be it the number of users/devices/threats is seeing an unprecedented growth and there is an urgent need to cordon the perimeter from risks especially since attacks(DoS, RPL etc) can have devastating consequences especially in critical IoT environments like health care, transportation and industrial applications.. This paper looks at Intrusion detection systems for IoT Networks as a supplement to the more traditional methods of computer security for C,I,A (Confidentiality, Integrity and Availability). The first section of the paper explored what is an IDS, what is a generic architecture for IDS in IoT environments and then looked at why traditional IDS systems might not work seamlessly in IoT environments. The next section surveyed various state of the art IDS systems for IoT use cases primarily using the more recent data sets, having IoT traces/traffic packets and summarized the findings along with taxonomy classification. The latter sections of this survey paper look at the pitfalls of using ML/ learning-based systems in this domain and critically evaluate the papers on several grounds. Finally, the challenges to be overcome in this domain are studied with some recommendations and potential future work.

# 9 Acknowledgements

The author would like to express his gratitude to Dr. Clifford Neuman (Associate Professor, USC) for his invaluable support in reviewing the paper contents and providing feedback. Dr. Neuman's experience and encouragement have been instrumental in piquing my interest in the field of Computer Security.

## 10 About the Authors

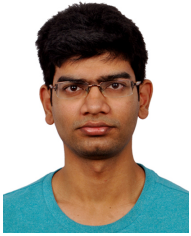

Vishal Karanam is a graduate student in Computer Science at the University of Southern California, Los Angeles. His current areas of interest are Security, Machine Learning and Distributed Systems. He's currently working with the Cloud Security team at Intel, Santa Clara. Vishal has previously worked with Amazon, Goldman Sachs and JP Morgan Chase in a variety of software roles and internships. He received his Bachelor's degree(Hons) in CS from the National Institute of Technology Calicut, India.